\documentclass[12pt]{article}

\newcommand{\m}{\mathrm}
\newcommand{\be}{\begin{equation}}
\newcommand{\ee}{\end{equation}}
\newcommand{\ba}{\begin{eqnarray}}
\newcommand{\ea}{\end{eqnarray}}

\usepackage{graphicx}
\usepackage{amssymb}
\usepackage{amsmath}
\usepackage[T1]{fontenc}
\usepackage[ansinew]{inputenc}
\usepackage[nosort]{cite}
\newcommand{\inbar}{\vrule height1.57ex width.4pt depth0pt}
\newcommand{\SW}{\relax{\hbox{$\ \inbar\kern-.285em{\rm S}$}}}

\usepackage{hyperref}
\usepackage{url}
\usepackage{color}

\definecolor{vividviolet}{rgb}{0.62, 0.0, 1.0}
\definecolor{purple(munsell)}{rgb}{0.62, 0.0, 0.77}
\definecolor{palatinateblue}{rgb}{0.15, 0.23, 0.89}
\hypersetup{linktoc=all,
    colorlinks, linkcolor={purple(munsell)},
    citecolor={palatinateblue}, urlcolor={palatinateblue}
}

\newcommand{\changeurlcolor}[1]{\hypersetup{urlcolor=#1}}

\begin{document}
\thispagestyle{empty}
\begin{center}

\null \vskip-1truecm \vskip2truecm

{\Large{\bf \textsf{On The Existence of a Holographic Description of the LHC Quark-Gluon Plasmas }}}


\vskip1truecm
\textbf{\textsf{Brett McInnes}}\\
{\footnotesize\textsf{Department of Mathematics, National
  University of Singapore, Singapore 119076 \\
{\tt Email: matmcinn@nus.edu.sg}}}\\

\vskip0.4truecm
\textbf{\textsf{Yen Chin Ong}}\\
{\footnotesize\textsf{Center for Astronomy and Astrophysics, Department of Physics and Astronomy,\\
Shanghai Jiao Tong University, Shanghai 200240, China}\\
{\tt Email: ongyenchin@sjtu.edu.cn}}\\

\end{center}

\vskip1truecm \centerline{\textsf{ABSTRACT}} \baselineskip=15pt
\medskip
Peripheral collisions of heavy ions can give rise to extremely intense magnetic fields. It has been suggested that these fields might invalidate the holographic description of the corresponding quark-gluon plasmas, assuming that these can be modelled by strongly coupled field theories. In the case of the plasmas produced in collisions at the RHIC facility (including in the beam energy scans), it is known how to deal with this problem: one has to take into account the large angular momenta generated in these plasmas, and the effects of the baryonic chemical potential. But this does not work for the plasmas produced in peripheral collisions at the LHC. However, these results neglect some (less significant) aspects of bulk physics; could it be that the problem is resolved by taking into account these lower-order effects? Here we use a bulk dilatonic field (fully compatible with boundary data, as well as with the asymptotically AdS character of the bulk geometry) as a model of these effects, and show that this is unlikely to be the solution. Thus, the existence of a consistent holographic description of the most extreme LHC plasmas remains open to question.

\newpage
\addtocounter{section}{1}
\section* {\large{\textsf{1. The Quark-Gluon Plasma in the LHC}}}
A basic question \cite{kn:raam} in the study of the gauge-gravity duality is this: which field theories have a gravity dual? In the case of applications to actual strongly coupled systems such as the Quark-Gluon Plasma \cite{kn:youngman,kn:gubser,kn:janik,kn:nat,kn:chessch}, this question becomes: does every \emph{realistic} strongly coupled system have such a dual? To settle this, one needs to examine the most extreme cases. The most extreme strongly-coupled systems currently accessible to experiment are probably (see below) the plasmas produced by collisions of heavy ions at the LHC \cite{kn:armesto,kn:aliceenergy}; so one needs to consider whether holography works in this case.

In \cite{kn:88} we adduced evidence \emph{suggesting that it does not}. The problem is a very fundamental one: it appears that the purported gravity dual in some cases does not exist when one attempts to interpret it (as one ultimately must \cite{kn:mateos}) as a string-theoretic system.

The situation may be briefly explained as follows. Ferrari and co-workers have shown \cite{kn:ferrari1,kn:ferrari2,kn:ferrari3,kn:ferrari4} that, simply for reasons of internal mathematical consistency, a string-theoretic bulk spacetime with a holographic dual must satisfy certain fundamental relations between the Euclidean spacetime action and the action of probes (such as branes). This has been explicitly confirmed in a large number of concrete cases \cite{kn:ferrari4}.

A specific example of such a relation is as follows: for \emph{every} $d\,-\,$dimensional hypersurface $\Sigma$ embedded in, and homologous to the conformal boundary of, a $(d+1)\,-\,$dimensional (Euclidean) bulk, the area $A(\Sigma)$ and the volume $V(M_{\Sigma})$ enclosed by $\Sigma$ are required to satisfy the inequality
\begin{equation}\label{A}
\mathfrak{S^{\m{E}}}\;\equiv\;A(\Sigma)\;- \;{d \over L}V(M_{\Sigma}) \;\geqslant \;0,
\end{equation}
where $L$ denotes the asymptotic AdS curvature scale, and the superscript ``E'' represents ``Euclidean''\footnote{The corresponding Lorentzian physics has been discussed elsewhere \cite{kn:84}. For the sake of clarity, in the present work we focus exclusively on the Euclidean case. Note that we are concerned here with black holes having flat, \emph{planar} event horizons (a zero superscript is used to remind us of this), so $\Sigma$ should be interpreted as a finite domain in such a plane. Since this domain can be chosen arbitrarily, there is an overall scale ambiguity in $\mathfrak{S^{\m{E}}}$, meaning that the scale on the vertical axes in all of our diagrams can be chosen at our convenience and has no physical significance. Notice however that $\mathfrak{S^{\m{E}}}$ must of course vanish at the ``centre'' of the Euclidean bulk, that is, at the Euclidean version of the event horizon; so there is no translational ambiguity along the vertical axis.}.

It is known \cite{kn:ferrari3} that this condition is satisfied by many candidate bulk geometries, including some very complicated ones such as Euclidean AdS-Kerr geometry. But quasi-realistic cases in which it is apparently \emph{not} satisfied are also known. In particular, the enormous \emph{magnetic fields} generated in the plasmas formed in some peripheral heavy-ion collisions \cite{kn:skokov,kn:tuchin,kn:magnet,kn:review,kn:hattori} are described by dual spacetimes in which ---$\,$ in the most extreme cases ---$\,$ the inequality (\ref{A}) is (seemingly) violated, as follows.

It was shown\footnote{The assumption in \cite{kn:82} is that the baryonic chemical potential is negligible; this is a reasonable approximation for the highest temperature plasmas at the RHIC, and an excellent one for the LHC plasmas, so we will maintain it throughout the present work.} in \cite{kn:82} that (\ref{A}) can be translated holographically (through the usual bulk black hole construction) to a relation between the magnetic field $B$ experienced by the boundary field theory and its temperature $T$: in natural units,
\begin{equation}\label{B}
B\;\leqslant \;2\pi^{3/2}T^2\;\approx \; 11.14 \times T^2.
\end{equation}
This is the holographic dual of the inequality (\ref{A}), in this specific case (in which the bulk black hole is described by a Euclidean asymptotically AdS magnetic Reissner-Nordstr\"om metric $g^E(\m{AdSP^*RN^{0})}$, given below, characterised by only two parameters, the magnetic parameter $P^*$ and the mass parameter $M^*$).

Recent analyses (see for example \cite{kn:fivetimes,kn:bzdak,kn:shipu,kn:holliday}) suggest that the magnetic fields encountered in some peripheral collisions may be much higher than previously thought: as high as $eB \approx 10 \times m_{\pi}^2$ (where $m_{\pi}$ is the conventional pion mass), or $B \approx 16.64$ fm$^{-2}$, even in RHIC collisions, for which the right side of (\ref{B}) is $\approx 13.97$ fm$^{-2}$ (with $T \approx 220$ MeV). Thus (\ref{B}) is violated in this case; and it is violated still more clearly in the corresponding LHC collisions. It appears, then, that we have a concrete physical system with a purported dual spacetime that violates (\ref{A}), and which is consequently mathematically inconsistent within string theory.

One can see this explicitly for these data, in Figure 1 (the vertical scale having been chosen for convenience, as explained above):
\begin{figure}[!h]
\centering
\includegraphics[width=0.65\textwidth]{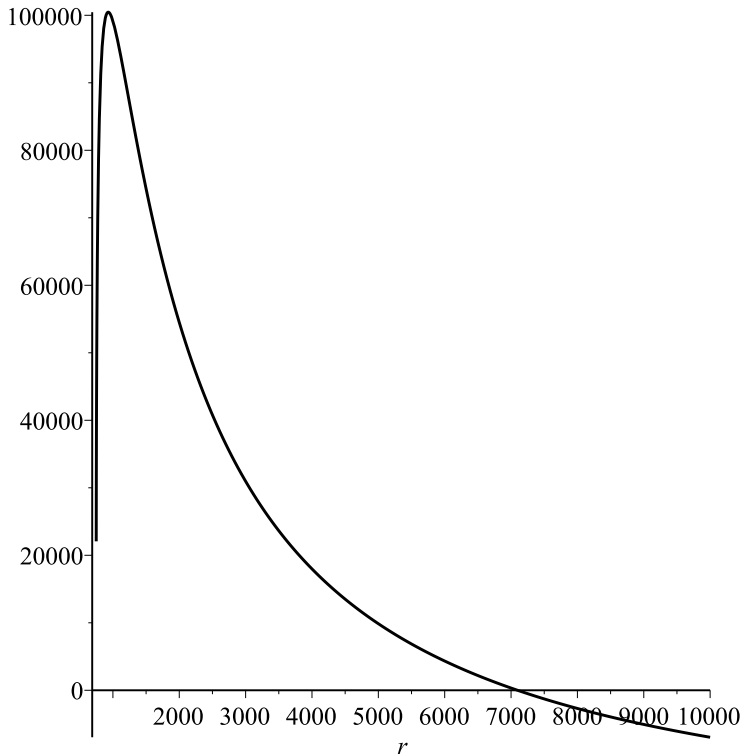}
\caption{$\mathfrak{S^{\m{E}}}(\m{AdSP^*RN^{0}_{4}})(r)$, $T \approx 220$ MeV, $eB \approx 10 \times m_{\pi}^2$.}
\end{figure}
Evidently $\mathfrak{S^{\m{E}}}(\m{AdSP^*RN^{0}_{4}})(r)$ (where $r$ is the radial black hole coordinate) is indeed negative for some values of $r$ in this case.

In \cite{kn:88} we argued however that it is not physically reasonable to consider magnetic fields in this situation without also considering the \emph{huge angular momentum densities} which also arise in peripheral collisions \cite{kn:liang,kn:bec,kn:huang,kn:KelvinHelm,kn:viscous,kn:csernairecent1,kn:csernairecent2,kn:nagy,kn:nacs,kn:yin,kn:deng,kn:vortical}: for the angular momentum is associated with the very mechanism (the internal motion of the plasma) that gives rise to the magnetic fields. This is directly relevant in a holographic context, because the geometry of a black hole spacetime is influenced by its angular momentum parameter. Since the inequality (\ref{A}) obviously depends on the bulk geometry, the presence of large angular momenta in the dual spacetime can have a bearing on the question as to whether it is satisfied. (In other words, in the presence of angular momentum, (\ref{A}) is no longer equivalent to (\ref{B}), but rather to some (much) more complicated inequality generalizing (\ref{B}).)

For the case of the RHIC plasmas, we found in \cite{kn:88} that including the shearing angular momentum generated by a peripheral collision (by generalizing the bulk metric to a certain Pleba\'nski--Demia\'nski metric \cite{kn:plebdem,kn:grifpod}, see \cite{kn:shear}) has a dramatic effect. Without it, as we have seen, (\ref{A}) is probably violated, but, with the inclusion of even a small amount of angular momentum, (\ref{A}) is immediately satisfied. Thus, a simple observation ---$\,$ that, in the aftermath of certain peripheral heavy-ion collisions, large magnetic fields are always accompanied by similarly large angular momenta ---$\,$ resolves the problem: these plasmas do have a dual description, albeit one involving a much more intricate bulk geometry than is usually considered\footnote{We note in passing that we found that non-negligible values of the baryonic chemical potential likewise resolve the problem; but this fact is of no use to us in the case of the LHC plasmas.}.

There is another way of looking at this result. The amount of angular momentum needed to restore (\ref{A}) for RHIC plasmas is in fact \emph{remarkably} small relative to the actual value: we estimate that the (specific) angular momentum\footnote{That is, angular momentum per unit of energy: in natural units, this has units of inverse energy or length.} in the case of RHIC plasmas with very large magnetic fields is in the range 50-75 fm, but values considerably smaller than \emph{one} fm ensure that (\ref{A}) holds: see Figure 2.
\begin{figure}[!h]
\centering
\includegraphics[width=0.65\textwidth]{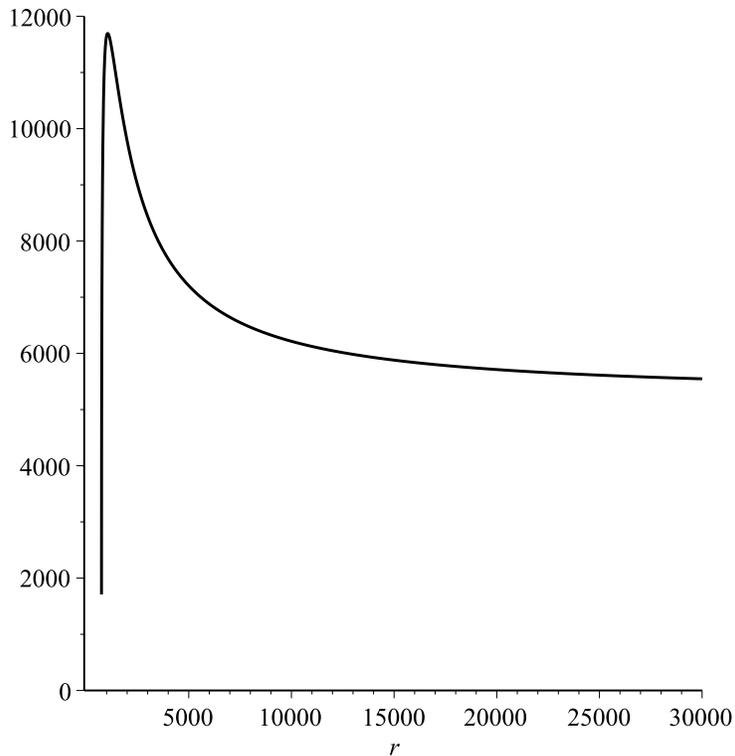}
\caption{As in Figure 1, but with a specific shearing angular momentum = 1 fm.}
\end{figure}
This means that neglecting angular momentum in this context (which, in fact, is precisely what is done in most of the literature) is not only physically unreasonable ---$\,$ it amounts to a kind of ``fine-tuning''. If one had begun with a \emph{generic} planar asymptotically AdS black hole geometry, with the specific angular momentum taking almost any non-zero value, then the question of RHIC plasmas violating (\ref{A}) would not have arisen.

But the case of the LHC plasmas is very different. Here, the magnetic fields attained in certain peripheral collisions are even more gigantic, ranging (see for example\cite{kn:denghuang,kn:gergend}) up to $eB \approx 70 \times m_{\pi}^2$ ($\approx$ 1.3 GeV$^2$), or $B \approx 117$ fm$^{-2}$, for a plasma temperature around 300 MeV, considerably exceeding the right side of the inequality (\ref{B}) in this case ($\approx 25.8$ fm$^{-2}$); but the angular momentum densities here are also much larger than for the RHIC plasma, maximal values for the specific angular momentum being over 450 fm. One consequently expected that the problem could be resolved in this case too, in the same way as for the RHIC plasma. Surprisingly, however, we found in \cite{kn:88} that this is \emph{not} the case: in fact, $\mathfrak{S^{\m{E}}}$ continues to take negative values for the maximal realistic value of the specific angular momentum. In short, angular momentum does \emph{not} resolve the violation of the consistency condition (\ref{A}) in the case of the LHC plasmas.

There are two possibilities at this point.

$\bullet$ Some authors question whether the LHC plasmas are indeed strongly coupled: see for example the discussion in \cite{kn:noodles}. (Recently it has been suggested that the LHC plasmas do differ from their RHIC counterparts in ways that remain to be fully understood \cite{kn:quench}; it is conceivable that this is relevant here.) If they are not strongly coupled, then gauge-gravity duality should not be applied to them in the first place, and there is no issue.

$\bullet$ On the other hand, we saw that the problem in the case of the RHIC plasma arose because we were unwittingly using a ``fine-tuned'' bulk geometry, one in which the black hole angular momentum parameter was tuned to be almost exactly zero. Perhaps some other small deformation of the bulk geometry, one which nevertheless maintains both the values of the boundary temperature and magnetic field and the asymptotic AdS geometry, can allow us to save the string-theoretic consistency of the theory. \emph{In short, we need to ask whether (\ref{A}) is being violated in the LHC case because some additional parameter is being tuned to unreasonably small values.}

As long as there is doubt regarding the strength of the LHC plasma coupling, the second possibility should be excluded before the first is accepted. We therefore propose to investigate whether (\ref{A}) can be restored by means of some ``small'' perturbation of the bulk black hole geometry corresponding to a boundary field theory that models the extreme LHC plasmas.

Our strategy is as follows. We wish to modify the Euclidean asymptotically AdS magnetic Reissner-Nordstr\"om planar black hole metric $g^E(\m{AdSP^*RN^{0})}$, while fixing the temperature and the magnetic field, and not disturbing the asymptotically AdS character of the bulk spacetime; but we want to have full control over the resulting geometry, since otherwise one cannot be certain whether (\ref{A}) holds or is violated. Within string theory, this confines us, in practice, to considering a \emph{dilaton} $\varphi$ with an adjustable coupling\footnote{The coupling is through a term of the form $e^{-2\alpha \varphi}F^2$ in the Lagrangian, where $F^2$ is the usual square of the electromagnetic two-form; as $\alpha$ enters all of our calculations only through its square, we take it to be positive; \emph{a priori}, $\alpha$ can take any positive value.} to the magnetic field; in fact, to preserve the asymptotically AdS geometry, we are forced to consider a specific dilaton potential\footnote{For a thorough discussion of the physics of this potential, and also of a remarkable application, see \cite{kn:abdalla}. For a comprehensive general survey of dilaton potentials in string theory (including that of Gao and Zhang), see \cite{kn:gout}.}, discovered by Gao and Zhang \cite{kn:gz}, who also found the corresponding exact black hole solutions. These spacetimes (suitably generalised to include magnetic charge) coincide precisely with the asymptotically AdS magnetic Reissner-Nordstr\"om planar black hole geometry when the coupling $\alpha = 0,$ but are continuously deformed away from the latter as $\alpha$ increases.

Our objective is to use the Gao-Zhang black hole geometry, by adjusting $\alpha$, to determine whether a small deformation of $g^E(\m{AdSP^*RN^{0})}$ (with values of $P^*$ and $M^*$ corresponding to LHC data) can restore\footnote{The general question of the effect of the dilaton on string-theoretic consistency in the bulk was first investigated in \cite{kn:ong}.} (\ref{A}). In this concrete context, the correct definition of ``small'' becomes clearer: ``small'' should mean that $\alpha$ is ``small''. We will see later that it is possible to be much more precise about this ``smallness'', since there is in fact an upper bound on $\alpha$ for given values of the boundary temperature and magnetic field; also, it can be interpreted to mean that the energy density (at some distinguished location, which we take to be the event horizon) contributed by the dilaton should be small compared to the energy density at that location due to the magnetic field. (By the Einstein equations, this can easily be formulated in terms of the relative sizes of the contributions to the Ricci curvature.) If the distortion of the bulk is small in these two senses, and if the corresponding value of $\alpha$ is such that (\ref{A}) holds, then we will conclude that the apparent failure of holography in this application was a mere artifact of using an over-simplified bulk geometry.

We stress that we are \emph{not} claiming that the dilaton necessarily appears as part of the holographic description of the plasma on the boundary: we are merely using it to give a controllable distortion of the bulk, while keeping the boundary temperature and magnetic field fixed. \emph{This distortion is intended to be a simple proxy for other bulk effects that we have previously been neglecting}. Since we are (by construction) keeping the perturbation small and not allowing it to change the boundary parameters, we can hope to justify neglecting any other effect it might have at infinity.

We find that, with LHC data for temperature and magnetic field, the consistency condition (\ref{A}) cannot be restored by a dilatonic distortion that is ``small'' in the senses we will define. In short, our finding in \cite{kn:88}, that the internal consistency condition for string theory in this context is not satisfied by a purported bulk dual of LHC plasmas associated with large magnetic fields, appears to be robust: it is not due to any fine-tuning of the manner in which the dilaton deforms the bulk geometry. The conclusion is that, in the LHC case, holography will only work if one can find a physical justification (in terms of the boundary physics) for a very substantial deformation of the bulk geometry away from the standard asymptotically AdS magnetic Reissner-Nordstr\"om planar black hole geometry.

We begin with a description of the bulk spacetime.

\addtocounter{section}{1}
\section* {\large{\textsf{2. The Dilaton Bulk Geometry }}}
The simplest possible bulk geometry dual to a QGP-like boundary theory subjected to a strong magnetic field is described as follows. We take a (Euclidean) asymptotically AdS magnetic Reissner-Nordstr\"om metric $g^E(\m{AdSP^*RN^{0})}$ of the form
\begin{eqnarray}\label{C}
g^E(\m{AdSP^*RN^{0})} & = &  \Bigg[{r^2\over L^2}\;-\;{8\pi M^*\over r}+{4\pi P^{*2}\over r^2}\Bigg]\m{d}t^2\; \nonumber \\
& &  + \;{\m{d}r^2\over {\dfrac{r^2}{L^2}}\;-\;{\dfrac{8\pi M^*}{r}}+{\dfrac{4\pi P^{*2}}{r^2}}} \;+\;r^2\left[\m{d}\psi^2\;+\;\m{d}\zeta^2\right];
\end{eqnarray}
here $L$ is the asymptotic AdS curvature scale, and $M^*$ and $P^*$ are parameters (with units of length) such that, if $r_h$ denotes the value of the radial coordinate at the event horizon (which has the geometry of a \emph{flat plane}, so that the spatial geometry at infinity is flat), then, if $\ell_P$ is the bulk Planck length, $M^*/\ell_P^2r_h^2$ is the mass per unit horizon area, and $P^*/\ell_Pr_h^2$ is the magnetic charge per unit horizon area; $r$ and $t$ are the usual radial and ``time'' coordinates; and $\psi$ and $\zeta$ are dimensionless coordinates on the plane.

At infinity, after a conformal re-scaling of the form $r^2/L^2$, one finds that $t$ is indeed proper time there; while $\psi$ and $\zeta$ define the standard coordinates $x = L\psi$ and $z=L\zeta$ in the reaction plane of a heavy-ion collision. We are effectively constructing a dual for a field theory propagating on the flat spacetime defined by this reaction plane. (Notice that this procedure sets the length scale for the boundary field theory at $L$ ---$\,$ for example, if we were to compactify $\psi$ and $\zeta$, the field theory would be defined on a flat torus with volume determined by $L$.)

This is the context in which the inequality (\ref{A}) reduces to (\ref{B}). In \cite{kn:88} we deformed this geometry by allowing the black hole to take on angular momentum: it turns out that this alleviates the tendency of strong magnetic fields to violate these conditions. Here we wish to deform it in another way, by coupling the magnetic field to a dilaton $\varphi$ (with a coupling constant $\alpha$, as above), while maintaining an AdS asymptotic geometry\footnote{In view of our discussion above, we should really combine these two moves, that is, we should consider a version of the Gao-Zhang metrics with non-zero angular momentum. However, since angular momentum is completely dominated by the magnetic field in the LHC case, this would add (very great) complexity without substantially modifying our results.}.

These black holes were constructed in \cite{kn:gz}. They can be interpreted as Kaluza-Klein reductions of certain near-extremal black branes \cite{kn:gout}; here we will regard them as asymptotically AdS planar dilatonic Reissner-Nordstr\"om black holes with magnetic charge parameter $P^*$, mass parameter $M^*$, and with Euclidean metric
\begin{equation}\label{D}
g^E(\m{AdSdilP^*RN}^{0})=U(r)\m{d}t^2 + {\m{d}r^2\over U(r)} + [f(r)]^2 \left[\m{d}\psi^2\;+\;\m{d}\zeta^2\right],
\end{equation}
where the coordinates are as before and where
\begin{equation}\label{E}
U(r)=-\frac{8\pi M^*}{r}\left[1-\frac{(1+\alpha^2)P^{*2}}{2M^*r}\right]^{\frac{1-\alpha^2}{1+\alpha^2}} + \frac{r^2}{L^2} \left[1-\frac{(1+\alpha^2)P^{*2}}{2M^*r}\right]^{\frac{2\alpha^2}{1+\alpha^2}},
\end{equation}
and
\begin{equation}\label{F}
f(r)^2 = r^2\left(1-\frac{(1+\alpha^2)P^{*2}}{2M^*r}\right)^{\frac{2\alpha^2}{1+\alpha^2}}.
\end{equation}
One sees that the dilaton has a rather complicated effect on the metric.

The dilaton itself, $\varphi$, a dimensionless function of the radial coordinate only, is given by
\begin{equation}\label{FOX}
e^{2\alpha \varphi (r)}\;=\;\left(1-\frac{(1+\alpha^2)P^{*2}}{2M^*r}\right)^{\frac{2\alpha^2}{1+\alpha^2}}.
\end{equation}

The electromagnetic field two-form\footnote{These results are obtained straightforwardly from the analogous relations in \cite{kn:gz}, using electromagnetic duality (applying the Hodge star operator to obtain the magnetic field). Note that the usual dimensionless coefficient (which would have been $4\pi$ if the event horizon had been spherical) has been absorbed into the definition of $P^*$.} corresponding to this black hole can be expressed as follows:
\begin{equation}\label{G}
F\;=\;{P^*\over \ell_P}\left(1-\frac{(1+\alpha^2)P^{*2}}{2M^*r}\right)^{\frac{2\alpha^2}{1+\alpha^2}}\,\m{d}\psi \wedge \m{d}\zeta.
\end{equation}
Using this to define a field at infinity, we are free to choose the overall scaling of the potential one-form, and consequently the field two-form; we choose the scale to be consistent with the boundary field theory scale $L$ discussed earlier\footnote{This is described in \cite{kn:myers} as fixing the relative normalization of the gauge and gravity kinetic terms; see the discussion there. See also \cite{kn:hartkov}, where essentially the same scaling is used.}: we then have $F_{\infty} = {P^*\over L^3}\,\m{d}x \wedge \m{d}z$, since $x = L\psi, z=L\zeta$. We interpret this, in the usual manner \cite{kn:hartkov}, as the magnetic field experienced by the boundary field theory:
\begin{equation}\label{GAG}
B_{\infty}\;=\; {P^*\over L^3}.
\end{equation}

The Euclidean ``event horizon\footnote{Since magnetic charge is not complexified in passing to the Euclidean domain, this formula is in fact the same as its Lorentzian counterpart.}'', that is, the central point in the Euclidean ($t, r$) plane, is located at $r = r_h$, related to the other parameters by
\begin{equation}\label{H}
-\frac{8\pi M^*}{r_h}\left[1-\frac{(1+\alpha^2)P^{*2}}{2M^*r_h}\right]^{\frac{1-\alpha^2}{1+\alpha^2}} + \frac{r_h^2}{L^2} \left[1-\frac{(1+\alpha^2)P^{*2}}{2M^*r_h}\right]^{\frac{2\alpha^2}{1+\alpha^2}} = 0,
\end{equation}
and the Hawking temperature of this black hole, obtained from the Euclidean metric in the usual way, is given by
\begin{eqnarray}\label{I}
4\pi T_{\infty}&=&{8\pi M^*\over r_h^2}\left(1-\frac{(1+\alpha^2)P^{*2}}{2M^*r_h}\right)^{{1-\alpha^2\over 1+\alpha^2}}\;-\;{4\pi (1-\alpha^2)P^{*2}\over r_h^3}\left(1-\frac{(1+\alpha^2)P^{*2}}{2M^*r_h}\right)^{{-2\alpha^2\over 1+\alpha^2}}\;\nonumber \\ &
&+\;{2r_h\over L^2}\left(1-\frac{(1+\alpha^2)P^{*2}}{2M^*r_h}\right)^{{2\alpha^2\over 1+\alpha^2}}\;+\;{\alpha^2P^{*2}\over M^*L^2}\left(1-\frac{(1+\alpha^2)P^{*2}}{2M^*r_h}\right)^{{\alpha^2 - 1 \over 1+\alpha^2}}.
\end{eqnarray}
This will of course be interpreted as the temperature of the boundary field theory.

The function $\mathfrak{S^{\m{E}}}$ defined in (\ref{A}) is given in this case, up to a positive constant factor, by
\begin{eqnarray}\label{J}
\mathfrak{S^{\m{E}}}(\m{AdSdilP^*RN^{0}})(r) & = &\frac{r^3}{L}\left[1-\frac{(1+\alpha^2)P^{*2}}{2M^*r}\right]^{\frac{3\alpha^2}{1+\alpha^2}}
\left[1-\frac{8\pi M^*L^2}{r^3}\left(1-\frac{(1+\alpha^2)P^{*2}}{2M^*r}\right)^{\frac{1-3\alpha^2}{1+\alpha^2}}\right]^{\frac{1}{2}}\nonumber \\ & &-\frac{3}{L}\int_{r_h}^r s^2\left[1-\frac{(1+\alpha^2)P^{*2}}{2M^*s}\right]^{\frac{2\alpha^2}{1+\alpha^2}} \m{d}s;
\end{eqnarray}
the integral can be expressed exactly in terms of a hypergeometric function\footnote{One finds\begin{equation}
\int_{r_h}^r s^2\left[1-\frac{(1+\alpha^2)P^{*2}}{2M^*s}\right]^{\frac{2\alpha^2}{1+\alpha^2}} \m{d}s \;=\;
_2F_1\left(4, 1;\frac{2\alpha^2+4}{1+\alpha^2};{2M^*r\over (1+\alpha^2)P^{*2}}\right) \;-\; _2F_1\left(4, 1;\frac{2\alpha^2+4}{1+\alpha^2};{2M^*r_h\over (1+\alpha^2)P^{*2}}\right).\nonumber\end{equation}} if one so desires.

Our objective is to determine, for values of $T_{\infty}$ and $B_{\infty}$ actually encountered in heavy ion collisions at the LHC, the circumstances under which this function is everywhere non-negative: that is, the conditions required for the supposed bulk dual actually to exist consistently within string theory.

The procedure is as follows: prescribe values for $\alpha$, $T_{\infty}$, $L$, and $B_{\infty}$ (so that, by equation (\ref{G}), $P^*$ is known). Then, \emph{if possible}, solve equations (\ref{H}) and (\ref{I}) as simultaneous equations for $M^*$ and $r_h$. We now have fixed all of the parameters in equation (\ref{J}), so we are in a position to determine whether this function is ever negative.

In practice, of course, this programme can only be fully carried out numerically. The results are surprising in several ways, and we organise them so as to make the overall structure as clear as possible.

\addtocounter{section}{1}
\section* {\large{\textsf{3. Varying the Dilaton-Magnetic Coupling }}}
Throughout this section, we fix values of $T_{\infty}$ and $B_{\infty}$ arising, for favourable values of the impact and other parameters, in plasmas formed by lead-lead collisions at the LHC: these we call the ``LHC data''. The specific values we use were discussed earlier: $T_{\infty} \approx 300$ MeV, $eB_{\infty} \approx 70 \times m_{\pi}^2$; we take $L$ to be a characteristic length scale of the plasma sample, $L \approx $ 10 fm. With this understanding, the only variable here is $\alpha$. Let us now see what a mainly (but not exclusively) numerical investigation reveals.
\subsubsection*{{\textsf{3.1. There is an upper bound on $\alpha$.}}}
When $\alpha = 0$, then of course (\ref{A}) is violated with LHC data; consequently it is violated also when $\alpha$ is extremely small. We therefore need to examine what happens as $\alpha$ is gradually increased.

In fact, as one adjusts $\alpha$ to larger values, one encounters something unexpected: a numerical investigation (supported by graphs of $M^*$ as a function of $r_h$, defined by equations (\ref{H}) and (\ref{I})) shows that, if $\alpha$ larger than a certain value, then equations (\ref{H}) and (\ref{I}) \emph{do not have any real solutions} for fixed values of $T_{\infty}$ and $B_{\infty}$.

We can describe this situation in the following very striking manner: the coupling between a very intense magnetic field and the dilaton can have such a strong effect on the physics of the bulk black hole that \emph{it cannot attain LHC temperatures}. Such a dramatic effect surely indicates that the magnetic-dilatonic coupling is \emph{strong}. This provides us with a natural definition of ``small'' values of $\alpha$.

In fact, for LHC data, the upper bound on $\alpha$ so imposed is smaller than one might have expected. If we define $\alpha^+(T_{\infty}$, $B_{\infty})$ as the value of $\alpha$ such that, for those values of $T_{\infty}$ and $B_{\infty}$, all values $\alpha > \alpha^+(T_{\infty}$, $B_{\infty})$ lead to a system given in equations (\ref{H}) and (\ref{I}) with no real solutions, then we find numerically that
\begin{equation}\label{K}
\alpha^+(T_{\infty} = 300\, \m{MeV}, eB_{\infty} = 70 \times m_{\pi}^2) \approx 0.605.
\end{equation}
In short, the dilaton-magnetic coupling has to be smaller than about 0.6 for our programme even to get off the ground; the range of $\alpha$ values available to us is in fact extremely narrow. To put it another way: in the context of our problem here, values of $\alpha$ comparable to $\alpha^+(T_{\infty} = 300\, \m{MeV}, eB_{\infty} = 70 \times m_{\pi}^2)$ have such a remarkable effect that we can justly claim that ``small'' $\alpha$ should be interpreted as meaning, ``small compared to $\alpha^+(T_{\infty} = 300\, \m{MeV}, eB_{\infty} = 70 \times m_{\pi}^2)$''. In concrete terms, we can take it that ``small'' $\alpha$ refers to values around an order of magnitude smaller than $\alpha^+(T_{\infty} = 300\, \m{MeV}, eB_{\infty} = 70 \times m_{\pi}^2)$: say, $\alpha \leqslant 0.1$.

We see that it is far from obvious that the dilaton can restore condition (\ref{A}): the danger is that values of $\alpha$ that are small in this sense may be too small to do so. We now consider this.
\subsubsection*{{\textsf{3.2. String Theory Imposes a Lower Bound on $\alpha$}}}
It is known \cite{kn:ong} that the dilaton does help to restore (\ref{A}): that is, it has the same sort of effect as angular momentum (and the opposite effect to that of magnetism). All that is required is that $\alpha$ be sufficiently large; but there are two senses in which this statement holds, as follows.

First, there is a critical value of $\alpha$, let us call it $\alpha_{\mathcal{C}}$, such that (\ref{A}) holds for \emph{any} $\alpha \geqslant \alpha_{\mathcal{C}}$, provided only that (\ref{H}) and (\ref{I}) have real solutions for $M^*$ and $r_h$; this statement is otherwise \emph{independent} of the values of $T_{\infty}$ and $B_{\infty}$. Second, if we are willing to accept a bound that does depend on $T_{\infty}$ and $B_{\infty}$ (for example, if we fix them at their extreme values in the LHC collisions), then we can find a lower bound. We consider these two types of bound in turn.
\subsubsection*{{\textsf{3.2.1. A General Lower Bound}}}

The quantity $\mathfrak{S^{\m{E}}}$ is in general quite difficult to study analytically. However, we note that $U(r)$ can be factorized if $\alpha=1/\sqrt{3}$:
\begin{equation}
U(r) =\left[\frac{r^2}{L^2}-\frac{8\pi M^*}{r}\right] \left[1-\frac{(1+\alpha^2)P^{*2}}{2M^* r}\right]^{\frac{1}{2}} = \left[\frac{r^2}{L^2}-\frac{8\pi M^*}{r}\right] \left[1-\frac{2P^{*2}}{3M^* r}\right]^{\frac{1}{2}}.
\end{equation}
In fact, $U(r)$ can be factorized in arbitrary spacetime dimension $n \geqslant 4$, when $\alpha = (n-3)/\sqrt{n-1}$. Let us denote this value of $\alpha$, which is fixed by the spacetime dimension\footnote{The quantity $\mathfrak{S^{\m{E}}}$ grows asymptotically linearly in four dimensions \cite{kn:84}, and asymptotically logarithmically in five dimensions \cite{kn:ong}. It can be shown that in higher dimensions, it is asymptotically a constant.}, by $\alpha_n$. (In five dimensions, $\alpha_5 = 1$. See \cite{1101.5776}.) We claim that $\alpha_n$ is an upper bound for $\alpha_{\mathcal{C}}$, although the bound may not be sharp.

In four dimensions, with $\alpha_4=1/\sqrt{3}\approx 0.578$, we see that the Euclidean event horizon\footnote{It might seem strange that the event horizon does not depend on the value of the magnetic charge (if we do not fix the boundary parameters). However, this is purely due to the choice of coordinate: recall that $r$ is not an area coordinate. The same behavior can be seen in the much simpler case of an asymptotically flat electrically charged dilaton black hole, where the Lorentzian event horizon is always $r_h=2M$ regardless of the value of $Q$ \cite{GHS,G,GM}.} satisfies
\begin{equation}\label{fix}
r_h^3 = 8\pi M^* L^2,
\end{equation}
so that for any fixed value of the cosmological constant, varying $M$ is equivalent to varying $r_h$.
The other zero of $U(r)$,
\begin{equation}
r_c = \frac{2P^{*2}}{3M^*},
\end{equation}
corresponds to the inner (Cauchy) horizon of the black hole in the Lorentzian counterpart of the geometry. It plays no role in the Euclidean discussion.

\begin{figure}[!h]
\centering
\includegraphics[width=0.75\textwidth]{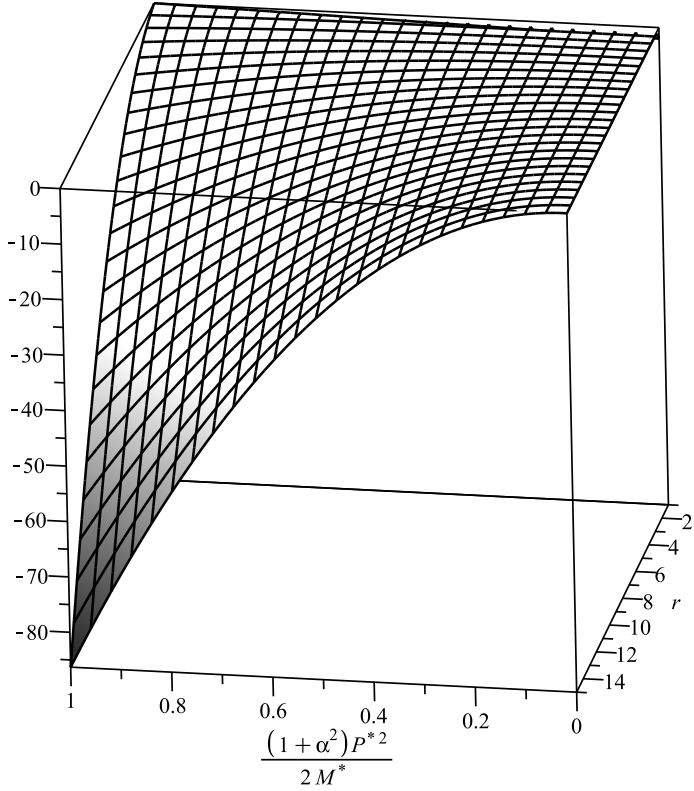}
\caption{The plot of the function $\mathcal{F}$ with the Euclidean horizon fixed at $r_h=1$, and $\alpha=\alpha_4=1/\sqrt{3}$.}\label{crit}
\end{figure}

In the following, we will keep $\alpha_4$ explicit without substituting in its numerical value, so that it is clear how the term enters the various quantities. We have
\begin{flalign}
\mathfrak{S^{\m{E}}}(\m{AdSdilP^*RN^{0}_{4}})(r) = &\frac{1}{L}\left\{r^{3}\left[1-\left(\frac{r_h}{r}\right)^{3}\right]^{\frac{1}{2}}\left[1-\left(\frac{(1+\alpha_4^2)P^{*2}}{2M^*r}\right)\right]^{\frac{3}{4}} \right. \notag \\ &\left. -3 \int_{r_h}^r s^{2} \left[1-\left(\frac{(1+\alpha_4^2)P^{*2}}{2M^*s}\right)\right]^{\frac{1}{2}} \m{d}s\right\}.
\end{flalign}
Our aim is to show that $\mathfrak{S^{\m{E}}} \geqslant 0$ for all values of $r \geqslant r_h$.
For dimension $n=4$, the derivative of $\mathfrak{S^{\m{E}}}$ is
\begin{equation}
\frac{\partial \mathfrak{S^{\m{E}}}}{\partial r} = -\frac{3}{4}\mathfrak{F}(r) \left[r^2 \sqrt{1-\frac{r_h^3}{r^3}}  \left(1-\frac{(1+\alpha_4^2)P^{*^2}}{2M^* r}\right)^{\frac{1}{4}}\right]^{-1},
\end{equation}
where
\begin{flalign}
\mathfrak{F}(r) = ~&4r^4\left[\left(1-\frac{(1+\alpha_4^2)P^{*^2}}{2M^* r}\right)^{\frac{3}{4}} \sqrt{1-\frac{r_h^3}{r^3}} -1\right] - r_h^3\left(\frac{(1+\alpha_4^2)P^{*^2}}{2M^*}\right) + 2r_h^3 r \\ \notag &+ 3\left(\frac{(1+\alpha_4^2)P^{*^2}}{2M^*}\right)r^3.
\end{flalign}
Since  $\mathfrak{S^{\m{E}}}$ vanishes on the Euclidean horizon $r_h$, it suffices to show that ${\partial \mathfrak{S^{\m{E}}}}/{\partial r} \geqslant 0$. That is, we want the function $\mathfrak{F}$ to be negative. This is clearly true at large enough values of $r$ since the $r^4$ term becomes dominant, and its coefficient is always negative. In Figure (\ref{crit}), we plot the function $\mathfrak{F}(r)$ by varying the value of ${(1+\alpha_4^2)P^{*2}}/{2M^*}$, but keeping $r_h$ fixed at unity. Since fixing $r_h$ is equivalent to fixing $M$, this means we are varying the magnetic charge. Changing the values of $r_h$ does not change the qualitative feature of the plot, namely that the function is always non-positive. From the plot, we see that the graph gets very close to zero as $P^{*2}$ approaches zero. Hence, one might be concerned that the value of the function $\mathfrak{F}$ is actually positive at some points, just that it is not visible at the resolution of the plot.

To see that this does not happen, we note that in the limit of small $P^{*2}$, the dominant terms are
\begin{equation}
\mathfrak{F}(r) \sim 4r^4\left(\sqrt{1-\frac{r_h^3}{r^3}}-1\right) + 2r_h^3 r = 4r^4\left(1-\frac{1}{2}\left(\frac{r_h^3}{r^3}\right) - \frac{1}{8}\left(\frac{r_h^3}{r^3}\right)^2 - O\left(\frac{r_h^3}{r^3}\right)^3\right) + 2r_h^3 r.
\end{equation}
That is,
\begin{equation}
\mathfrak{F}(r) \sim - \frac{1}{8}\left(\frac{r_h^3}{r^3}\right)^2 - O\left(\frac{r_h^3}{r^3}\right)^3 < 0,
\end{equation}
so indeed the function remains negative when $P^* \to 0$. This is not really surprising, since the dilaton is coupled to the magnetic field in such a way that if $P^*\to 0$, then $\varphi \to 0$, as can be seen from Eq.(\ref{FOX}). That is, in this limit the black hole has neither magnetic charge nor scalar field; it simply reduces to a neutral planar black hole, for which the associated $\mathfrak{S^{\m{E}}}$ is always non-negative.

Therefore, as we claimed, $\mathfrak{S^{\m{E}}}$ is non-negative for $\alpha \geqslant \alpha_4 = 1/\sqrt{3} \geqslant \alpha_\mathcal{C} $.  Note that we have only showed that $\alpha_4$ is an upper bound for $\alpha_\mathcal{C}$, the latter may take a lower value. However, we suspect that $\alpha_\mathcal{C}$ is not too far from $\alpha_4$, see below. What is surprising, however, is that $\alpha_4$ (and in general $\alpha_n$) is \emph{completely fixed by the spacetime dimension}. In particular, it does not depend on any of the other black hole parameters (namely the mass and magnetic charge density). This shows that with large enough $\alpha$, the bulk geometry will be sufficiently deformed such that $\mathfrak{S^{\m{E}}}\geqslant 0$, regardless of the temperature $T_{\infty}$ and the magnetic field strength $B_{\infty}$ at the boundary.

We emphasize here that fixing the value of $r_h$ here is physically very different from a similar ``horizon-fixing'' procedure that was performed in \cite{kn:84}. There, an electrically charged, Lorentzian version of the geometry was investigated (recall that mathematically this is in fact equivalent to our problem, since magnetic charge is not complexified under Wick rotation, and the Lorentzian geometry\footnote{Here, and throughout this work, we are referring to the geometry in the Einstein frame. The geometry as seen by a string is a conformally related string frame metric, in which a Lorentzian magnetically charged black hole is quite different from an electrically charged one. Essentially, this is because the dilaton field changes sign under electric-magnetic duality transformation.
See, e.g., \cite{9210119}.} is invariant under interchanging $Q^*$ and $P^*$). Indeed, by fixing $r_h$ at various values as was done in \cite{kn:84}, one could evaluate $\mathfrak{S^{\m{E}}}$ to see at which value of $\alpha=\bar{\alpha}$ the quantity $\mathfrak{S^{\m{E}}}$ becomes positive (at all values of $r$) for $\alpha > \bar{\alpha}$. However, $\bar{\alpha}$ \emph{depends} on the choice of the value of $r_h$. In other words, fixing the value of $r_h$ renders the result only qualitatively correct. We proposed in \cite{kn:84} that we should take the smallest value of $\bar{\alpha}$ to be the value of the critical value of $\alpha_\mathcal{C}$, which we numerically estimated to be $\alpha_\mathcal{C} \approx 0.53$. This estimate is indeed not too far from $\alpha_4=1/\sqrt{3}\approx 0.578$.

The reason why fixing $r_h$ in the manner of \cite{kn:84} leads to choice dependent $\mathfrak{S^{\m{E}}}$ is because this procedure \emph{fixes} the relation between the charge and the mass of the black hole. To see this, let us consider an asymptotically flat magnetically charged Reissner-Nordstr\"om black hole, whose horizon satisfies $r_h = M + \sqrt{M^2 - (P^2/4\pi)}$. Fixing $r_h=1$, for example, means that $P^2=4\pi(2M-1)$.

In this work, while we have fixed the value of $r_h$ in the analysis above, the result is \emph{independent} of the choice of $r_h$. This is because for $\alpha=\alpha_n$, the function $U(r)$ factorizes in such a way that choosing a value of $r_h$ is the same as, via Eq.(\ref{fix}), fixing the mass only. In other words, mass and charge remain as separate parameters that can be tuned separately.

Having shown that string theory imposes a general lower bound on $\alpha$, let us now turn to a bound which applies specifically to the LHC data.
\subsubsection*{{\textsf{3.2.2. A Lower Bound Given LHC Data}}}

Since $\alpha_{\mathcal{C}} < \alpha^+(T_{\infty} = 300\, \m{MeV}, eB_{\infty} = 70 \times m_{\pi}^2),$ we see that it is certainly possible to use the dilaton to solve our problem: with a value of the coupling in the narrow band between these limits, (\ref{A}) holds even with LHC data for the temperature and the magnetic field. Unfortunately, however, $\alpha_{\mathcal{C}}$ is by no means small, in the sense we defined above.

While values of $\alpha$ above $\alpha_{\mathcal{C}}$ enforce the consistency condition, this does not mean that lower values cannot perform this service, though the range of such $\alpha$ \emph{will} depend on the specific values of $(T_{\infty}$, $B_{\infty})$. Clearly we should focus on a value of $\alpha$ which is as small as possible for given $(T_{\infty}$, $B_{\infty})$: that is, we should try to determine $\alpha^-(T_{\infty}$, $B_{\infty}),$ defined to be such that (\ref{A}) is violated for any $\alpha < \alpha^-(T_{\infty}$, $B_{\infty}).$ For LHC data, we have found numerically that this quantity is given as follows:
\begin{equation}\label{L}
\alpha^-(T_{\infty} = 300\, \m{MeV}, eB_{\infty} = 70 \times m_{\pi}^2) \approx 0.284.
\end{equation}
One can see this in the graph of $\mathfrak{S^{\m{E}}}(\m{AdSdilP^*RN^{0}})(r)$ for these parameter values\footnote{A close examination of the graph shows that it does not actually touch the horizontal axis; so in fact $\alpha^-(T_{\infty} = 300\, \m{MeV}, eB_{\infty} = 70 \times m_{\pi}^2)$ must be slightly smaller than 0.284.}: see Figure 4.
\begin{figure}[!h]
\centering
\includegraphics[width=0.65\textwidth]{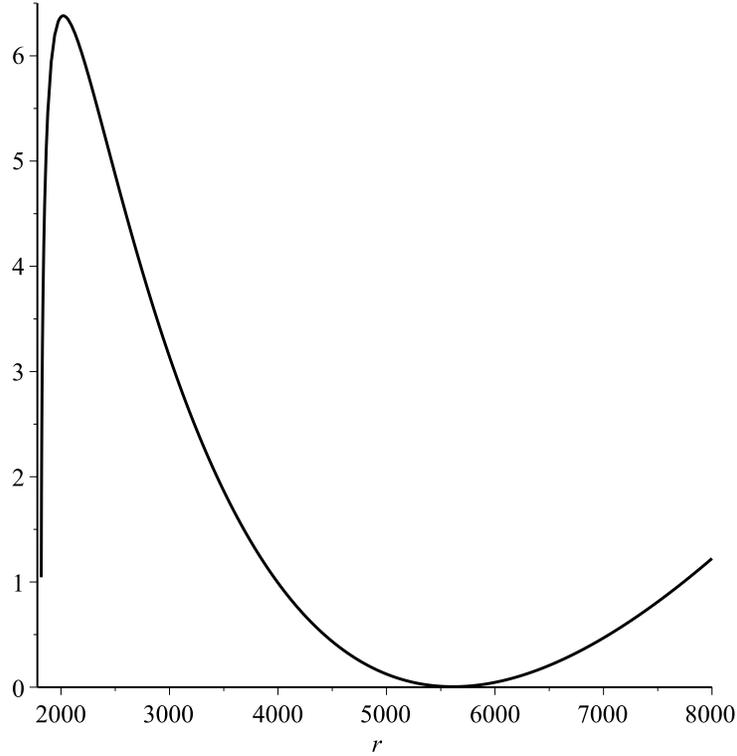}
\caption{$\mathfrak{S^{\m{E}}}(\m{AdSdilP^*RN^{0}})(r)$, LHC data, $\alpha = 0.284$}
\end{figure}

In summary, then, the dilaton can resolve the apparent conflict between the LHC data and the bulk consistency condition (\ref{A}); but it can only do so by means of a magnetic-dilatonic coupling that \emph{does not correspond to a small perturbation} of the bulk geometry, the minimal value of the coupling being about half of the maximal possible value.

\subsubsection*{{\textsf{3.3. Relative Energy of the Dilaton}}}
It may seem odd to describe $\alpha \approx 0.3$ as a ``large'' value for the dilaton-magnetic coupling, so let us investigate more directly the effect of such a field on the bulk geometry. In particular, we should consider in more detail the black hole geometry for the relevant values of $\alpha$ (by which we mean typical values between $\alpha^-(T_{\infty} = 300\, \m{MeV}, eB_{\infty} = 70 \times m_{\pi}^2)$ and $\alpha^+(T_{\infty} = 300\, \m{MeV}, eB_{\infty} = 70 \times m_{\pi}^2)$: say around $\alpha =$ 0.4 --- 0.5). For this purpose, it is useful to compare the relative contributions of the dilaton and the magnetic fields to the energy density (as measured by a Killing observer) at a distinguished location, the event horizon. The two are of course related, in the sense that both depend on $\alpha$. But they depend on it in different ways, so the outcome is unclear.

We begin with a computation of the energy density of the magnetic field, evaluated at the event horizon. From equations (\ref{D}) and (\ref{G}) we have
\begin{equation}\label{M}
F^{\psi\zeta}F_{\psi\zeta}\;=\;f(r)^{-4}{P^{*2}\over \ell_P^2}\left(1-\frac{(1+\alpha^2)P^{*2}}{2M^*r}\right)^{\frac{4\alpha^2}{1+\alpha^2}}.
\end{equation}
Recalling that the coupling of the magnetic and dilatonic fields is of the form $e^{-2\alpha \varphi}F^2$, and using the usual form of the electromagnetic stress-energy-momentum tensor together with equations (\ref{F}) and (\ref{FOX}), we have finally, at the event horizon,
\begin{equation}\label{N}
\rho_B(r_h)\;=\;{P^{*2}\over 2\ell_P^2r_h^4}\left(1-\frac{(1+\alpha^2)P^{*2}}{2M^*r_h}\right)^{\frac{-2\alpha^2}{1+\alpha^2}}.
\end{equation}
Using the Einstein equation, one can think of this as a measure of the extent to which the presence of a magnetic field distorts the geometry away from that of a simple (planar) AdS-Schwarzschild black hole.

The dilaton stress-energy-momentum tensor is just the usual one, and so, since $\varphi(r)$ depends only on $r$,
\begin{equation}\label{O}
\rho^{AdS}_{\varphi}\;=\;{1\over 2\ell_P^2}\partial^{\,r}\varphi\partial_r\varphi\,+\,V(\varphi),
\end{equation}
where $V(\varphi)$ is the dilaton potential, given in this case \cite{kn:gz} by
\begin{equation}\label{P}
V(\varphi)\;=\;{- 1\over 8\pi \ell_P^2L^2}{1\over (1+\alpha^2)^2}\left[\alpha^2\left(3\alpha^2-1\right)e^{-2\varphi/\alpha}\,+\,\left(3-\alpha^2\right)
e^{2\alpha \varphi}\,+\,8\alpha^2e^{\alpha \varphi - \left(\varphi/\alpha\right)}\right].
\end{equation}

Notice that $V(\varphi) \rightarrow -\, 3/\left(8\pi \ell_P^2L^2\right)$ when \emph{either} $\alpha \rightarrow 0$ \emph{or} $r \rightarrow \infty$ (see equation (\ref{FOX})); as this is just the ``energy density'' corresponding to a negative cosmological constant, we are reminded that we are indeed dealing with an asymptotically AdS geometry here. This also reminds us that there is an AdS energy density included in $V(\varphi)$, which should be subtracted if we are interested in analysing the purely dilatonic contribution to the energy density; that is why we denoted the energy density above by $\rho^{AdS}_{\varphi}$, to indicate that this subtraction has not yet been performed. Notice too that $\partial^{\,r}\varphi = U(r)\partial_r\varphi$ (see equation (\ref{D})), which vanishes at the horizon; so to evaluate $\rho^{AdS}_{\varphi}$ in equation (\ref{O}) at the horizon we need only to consider the value of $V(\varphi)$ there.

Using (\ref{FOX}) and subtracting the background AdS energy density $- \,3/\left(8\pi \ell_P^2L^2\right)$ we obtain, for the purely dilatonic contribution to the energy density at the event horizon,
\begin{eqnarray}\label{Q}
\rho_{\varphi}(r_h) & = & {- 1\over 8\pi \ell_P^2L^2}\left\{{1\over (1+\alpha^2)^2}\left[\alpha^2\left(3\alpha^2-1\right)\left(1-\frac{(1+\alpha^2)P^{*2}}{2M^*r_h}\right)^{\frac{-\,2}{1+\alpha^2}} \right. \right.\\
&& \left. \left. +\,\left(3-\alpha^2\right)
\left(1-\frac{(1+\alpha^2)P^{*2}}{2M^*r_h}\right)^{\frac{2\alpha^2}{1+\alpha^2}}
\,+\,8\alpha^2\left(1-\frac{(1+\alpha^2)P^{*2}}{2M^*r_h}\right)^{\frac{\alpha^2 - 1}{1+\alpha^2}}\right]\;-\;3\right\}. \nonumber
\end{eqnarray}
From equation (\ref{N}) we therefore have finally
\begin{eqnarray}
{|\rho_{\varphi}(r_h)|\over \rho_B(r_h)} & = & {r_h^4\over 4\pi P^{*2}L^2}\left(1-\frac{(1+\alpha^2)P^{*2}}{2M^*r_h}\right)^{\frac{2\alpha^2}{1+\alpha^2}}\left\{{1\over (1+\alpha^2)^2}\left[\alpha^2\left(3\alpha^2-1\right)\left(1-\frac{(1+\alpha^2)P^{*2}}{2M^*r_h}\right)^{\frac{-\,2}{1+\alpha^2}} \right. \right. \nonumber \\
 && \left. \left. +\,\left(3-\alpha^2\right)
\left(1-\frac{(1+\alpha^2)P^{*2}}{2M^*r_h}\right)^{\frac{2\alpha^2}{1+\alpha^2}}
\,+\,8\alpha^2\left(1-\frac{(1+\alpha^2)P^{*2}}{2M^*r_h}\right)^{\frac{\alpha^2 - 1}{1+\alpha^2}}\right]\;-\;3\right\}. \nonumber
\end{eqnarray}
Again, we can interpret this as a measure of the effect of the dilaton on the background bulk geometry, relative to the effect of the magnetic field.

This quantity is to be regarded as a function of the boundary data (temperature and magnetic field) together with $\alpha$. (Recall that $P^*$ is determined by $B_{\infty}$, and $M^*$ and $r_h$ are determined by solving equations (\ref{H}) and (\ref{I}) as simultaneous equations.) For values of $\alpha$ that we consider small, $\alpha \leqslant 0.1$, we find
\begin{equation}\label{R}
{|\rho_{\varphi}(r_h)|\over \rho_B(r_h)}\left(T_{\infty} = 300\, \m{MeV}, eB_{\infty} = 70 \times m_{\pi}^2, \alpha = 0.1\right) \approx 0.0076 .
\end{equation}
This may indeed be interpreted as meaning that the dilaton only causes a relatively small perturbation.

At the other extreme, if $\alpha$ is chosen to have its maximal possible value, around 0.605 as above, then one finds
\begin{equation}\label{S}
{|\rho_{\varphi}(r_h)|\over \rho_B(r_h)}\left(T_{\infty} = 300\, \m{MeV}, eB_{\infty} = 70 \times m_{\pi}^2, \alpha = 0.605\right) \approx 3.06,
\end{equation}
so the dilaton completely dominates the magnetic field, and our bulk geometry is far indeed from being a small perturbation away from the asymptotically AdS magnetic Reissner-Nordstr\"om planar black hole spacetime.

Thus we now have a more explicit formulation of our ``small'' and ``large'' values for $\alpha$: they correspond, respectively, to values of the relative dilaton/magnetic field energy density below 0.0076 and above 3. This seems reasonable.

For generic values of $\alpha$ capable of maintaining the consistency condition (\ref{A}), that is, typical values between $\alpha^-(T_{\infty} = 300\, \m{MeV}, eB_{\infty} = 70 \times m_{\pi}^2)$ and $\alpha^+(T_{\infty} = 300\, \m{MeV}, eB_{\infty} = 70 \times m_{\pi}^2)$, one has
\begin{equation}\label{T}
{|\rho_{\varphi}(r_h)|\over \rho_B(r_h)}\left(T_{\infty} = 300\, \m{MeV}, eB_{\infty} = 70 \times m_{\pi}^2, \alpha = 0.4 \;\textup{---}\; 0.5\right) \approx 0.200 \;\textup{---}\; 0.465;
\end{equation}
one would not say that such values have been fine-tuned to be near zero. This confirms our assessment that the violation of (\ref{A}) by LHC data is not due to fine-tuning.

\addtocounter{section}{1}
\section* {\large{\textsf{4. Conclusion}}}
It has been observed \cite{kn:barb} that the holographic techniques that work remarkably well when applied to the RHIC plasmas do not appear to work as well for LHC plasmas. This may well simply indicate that more elaborate holographic techniques \cite{kn:ficnar1,kn:ficnar2} are needed in the LHC case; but perhaps it indicates that there are fundamental obstacles to using holography in these extreme conditions.

In \cite{kn:88}, we found evidence for this second explanation, in terms of a violation of the fundamental string-theoretic consistency condition (\ref{A}); but the results were obtained with the simplest possible bulk geometry adequate to describe the dominant physical parameters of the boundary field theory, its temperature, magnetic field, and angular momentum density. That is, other possible bulk parameters were tuned to zero.

In this work, we have considered whether this tuning amounts to fine-tuning. Our conclusion is that it does not: the only way to rescue (\ref{A}) from the effects of strong magnetic fields is to distort the bulk with a dilaton which is strongly coupled to the bulk magnetic field. Unless one can find a justification for such a strong distortion in terms of QGP physics, it seems that a holographic description is not appropriate here.

The possible existence of such a justification should not be ruled out completely, since of course all known holographic models of the QGP are greatly over-simplified. For example, we have not considered the fact that the magnetic field and other plasma parameters are strongly time-dependent. While that in itself does not invalidate the usual approach using static black holes (since for example the consistency condition must hold at each instant of time), it is conceivable that there are non-trivial constraints on the dilaton coupling in the dynamical case; perhaps $\alpha$ does need to be ``large'', for physical reasons. However, we are unaware of any such effect. (Note that an attempt is made in \cite{kn:abdalla} to generalize the Gao-Zhang black holes to the dynamical case, by using a Vaidya-like geometry. There is no sign of a lower bound on $\alpha$ in that work.)

The alternative approach is to accept that asymptotic freedom will make itself felt at some point as collision energies increase: certainly one expects this to happen at future facilities \cite{kn:SppC,kn:FCC1,kn:FCC2}, potentially studying collision energies ranging up to 40 TeV. The QGP will then no longer be strongly coupled. Perhaps this is what we are beginning to see at the LHC.

It may be preferable to turn attention to the exciting prospects opened up by the various beam energy scan experiments currently under way or in preparation \cite{kn:STAR,kn:shine,kn:nica,kn:fair,kn:BEAM,kn:luo}, where the QGP is much more likely to be strongly coupled and where, as was explained in \cite{kn:88}, there is no difficulty in satisfying the consistency condition even in the presence of large magnetic fields (because the non-zero baryonic chemical potential tends to counteract the tendency of the magnetic field to violate it). A holographic approach may well prove fruitful in that region of the quark matter phase diagram: see for example \cite{kn:86,kn:87} and references therein.

\addtocounter{section}{1}
\section*{\large{\textsf{Acknowledgements}}}
BMc wishes to acknowledge helpful discussions with Jude and Cate McInnes; YCO wishes to thank NNSFC for support.


\begin{thebibliography}{18}

\bibitem{kn:raam}
Mark Van Raamsdonk, ``Lectures on Gravity and Entanglement'', \href{https://arxiv.org/abs/1609.00026}{[arXiv:1609.00026 [hep-th]]}.

\bibitem{kn:youngman}
Youngman Kim, Ik Jae Shin, Takuya Tsukioka, ``Holographic QCD: Past, Present, and Future'', {\changeurlcolor{vividviolet}\href{http://www.sciencedirect.com/science/article/pii/S0146641012001111}{Prog. Part. Nucl. Phys. \textbf{68} (2013) 55}}, \href{https://arxiv.org/abs/1205.4852}{[arXiv:1205.4852 [hep-ph]]}.

\bibitem{kn:gubser}
Oliver DeWolfe, Steven S. Gubser, Christopher Rosen, Derek Teaney, ``Heavy Ions and String Theory'',  {\changeurlcolor{vividviolet}\href{http://www.sciencedirect.com/science/article/pii/S0146641013001075}{Prog. Part. Nucl. Phys. \textbf{75} (2014) 86}}, \href{https://arxiv.org/abs/1304.7794}{[arXiv:1304.7794 [hep-th]]}.

\bibitem{kn:janik}
Romuald A. Janik, ``AdS/CFT and Applications'',  PoS EPS-HEP \textbf{2013} (2013) 141, \href{https://arxiv.org/abs/1311.3966}{[arXiv:1311.3966 [hep-ph]]}.

\bibitem{kn:nat}
Makoto Natsuume,	``AdS/CFT Duality User Guide'', {\changeurlcolor{vividviolet}\href{http://link.springer.com/book/10.1007/978-4-431-55441-7}{Lect. Notes Phys. \textbf{903} (2015) 1}}, \href{https://arxiv.org/abs/1409.3575}{[arXiv:1409.3575 [hep-th]]}.

\bibitem{kn:chessch}
Paul M. Chesler, Wilke van der Schee, ``Early Thermalization, Hydrodynamics and Energy Loss in AdS/CFT'', {\changeurlcolor{vividviolet}\href{http://www.worldscientific.com/doi/10.1142/S0218301315300118}{Int. J. Mod. Phys. E \textbf{24} (2015) 10, 1530011}}, \href{https://arxiv.org/abs/1501.04952}{[arXiv:1501.04952 [nucl-th]]}.

\bibitem{kn:armesto}
N\'estor Armesto, Enrico Scomparin, ``Heavy-Ion Collisions at the Large Hadron Collider: A Review of the Results from Run 1'', {\changeurlcolor{vividviolet}\href{http://link.springer.com/article/10.1140\%2Fepjp\%2Fi2016-16052-4}{Eur. Phys. J. Plus \textbf{131} (2016) 52}}, \href{https://arxiv.org/abs/1511.02151}{[arXiv:1511.02151 [nucl-ex]]}.


\bibitem{kn:aliceenergy}
ALICE Collaboration, ``Measurement of Transverse Energy at Midrapidity in Pb-Pb Collisions at $\sqrt{s_{NN}}$=2.76 TeV'', {\changeurlcolor{vividviolet}\href{http://journals.aps.org/prc/abstract/10.1103/PhysRevC.94.034903}{Phys. Rev. C \textbf{94} (2016) 034903}}, \href{https://arxiv.org/abs/1603.04775}{[arXiv:1603.04775 [nucl-ex]]}.


\bibitem{kn:88}
Brett McInnes, ``Field Theories Without a Holographic Dual'', to appear in Nucl. Phys. B, \href{https://arxiv.org/abs/1606.02425}{[arXiv:1606.02425 [hep-th]]}.

\bibitem{kn:mateos}
David Mateos, ``Gauge/String Duality Applied to Heavy Ion Collisions: Limitations, Insights and Prospects'', {\changeurlcolor{vividviolet}\href{http://iopscience.iop.org/article/10.1088/0954-3899/38/12/124030/meta;jsessionid=F6BCBEDB03C299AE55972F163B3E7AAC.c3.iopscience.cld.iop.org}{J. Phys. G \textbf{38} (2011) 124030}}, \href{https://arxiv.org/abs/1106.3295}{[arXiv:1106.3295 [hep-th]]}.

\bibitem{kn:ferrari1}
Frank Ferrari, ``Gauge Theories, D-Branes and Holography'', {\changeurlcolor{vividviolet}\href{http://www.sciencedirect.com/science/article/pii/S0550321314000182}{Nucl. Phys. B \textbf{880} (2014) 247}}, {\href{https://arxiv.org/abs/1310.6788}{[arXiv:1310.6788 [hep-th]]}}.

\bibitem{kn:ferrari2}
Frank Ferrari, ``D-Brane Probes in the Matrix Model'', {\changeurlcolor{vividviolet}\href{http://www.sciencedirect.com/science/article/pii/S0550321313006068}{Nucl. Phys. B \textbf{880} (2014) 290}}, \href{https://arxiv.org/abs/1311.4520}{[arXiv:1311.4520 [hep-th]]}.

\bibitem{kn:ferrari3}
Frank Ferrari, Antonin Rovai, ``Holography, Probe Branes and Isoperimetric Inequalities'', {\changeurlcolor{vividviolet}\href{http://www.sciencedirect.com/science/article/pii/S0370269315004268}{Phys. Lett. B \textbf{747} (2015) 212}}, \href{https://arxiv.org/abs/1411.1887}{[arXiv:1411.1887 [hep-th]]}.

\bibitem{kn:ferrari4}
Frank Ferrari, Antonin Rovai, ``Gravity and On-Shell Probe Actions'', {\changeurlcolor{vividviolet}\href{http://link.springer.com/article/10.1007\%2FJHEP08\%282016\%29047}{JHEP \textbf{08} (2016) 047}}, \href{https://arxiv.org/abs/1602.07177}{[arXiv:1602.07177 [hep-th]]}.

\bibitem{kn:84}
Brett McInnes, Yen Chin Ong, ``When Is Holography Consistent?'', {\changeurlcolor{vividviolet}\href{http://www.sciencedirect.com/science/article/pii/S0550321315002412}{Nucl. Phys. B \textbf{898} (2015) 197}}, \href{https://arxiv.org/abs/1504.07344}{[arXiv:1504.07344 [hep-th]]}.

\bibitem{kn:skokov}
Vladimir Skokov,  Alexey Yu. Illarionov, Viacheslav Toneev, ``Estimate of the Magnetic Field Strength in Heavy-Ion Collisions'', {\changeurlcolor{vividviolet}\href{http://www.worldscientific.com/doi/abs/10.1142/S0217751X09047570}{Int. J. Mod. Phys. A \textbf{24} (2009) 5925}}, \href{https://arxiv.org/abs/0907.1396}{[arXiv:0907.1396 [nucl-th]]}.

\bibitem{kn:tuchin}
Kirill Tuchin, ``Particle Production in Strong Electromagnetic Fields in Relativistic Heavy-Ion Collisions'', {\changeurlcolor{vividviolet}\href{https://www.hindawi.com/journals/ahep/2013/490495/}{Adv. High Energy Phys. \textbf{2013} (2013) 490495}}, \href{https://arxiv.org/abs/1301.0099}{[arXiv:1301.0099 [hep-ph]]}.

\bibitem{kn:magnet}
Xu-Guang Huang, ``Electromagnetic Fields and Anomalous Transports in Heavy-Ion Collisions --- A Pedagogical Review'', {\changeurlcolor{vividviolet}\href{http://iopscience.iop.org/article/10.1088/0034-4885/79/7/076302/meta}{Rep. Prog. Phys.\textbf{79} (2016) 076302}}, \href{https://arxiv.org/abs/1509.04073}{[arXiv:1509.04073 [nucl-th]]}.

\bibitem{kn:review}
Dmitri E. Kharzeev, Karl Landsteiner, Andreas Schmitt, Ho-Ung Yee (Editors), \emph{Strongly Interacting Matter in Magnetic Fields}, {\changeurlcolor{vividviolet}\href{http://link.springer.com/book/10.1007\%2F978-3-642-37305-3}{Lect. Notes Phys. \textbf{871} (2013) 1}}, \href{https://arxiv.org/abs/1211.6245}{[arXiv:1211.6245[hep-ph]]}.

\bibitem{kn:hattori}
Koichi Hattori, Xu-Guang Huang,
``Novel Quantum Phenomena Induced by Strong Magnetic Fields in Heavy-Ion Collisions'', \href{https://arxiv.org/abs/1609.00747}{[arXiv:1609.00747 [nucl-th]]}.

\bibitem{kn:82}
Brett McInnes, ``A Holographic Bound on Cosmic Magnetic Fields'', {\changeurlcolor{vividviolet}\href{http://www.sciencedirect.com/science/article/pii/S0550321315000036}{Nucl. Phys. B \textbf{892} (2015) 49}}, \href{https://arxiv.org/abs/1409.3663}{[arXiv:1409.3663 [hep-th]]}.

\bibitem{kn:fivetimes}
Vadim Voronyuk, Viacheslav D. Toneev, Wolfgang Cassing, Elena L. Bratkovskaya, Volodymyr P. Konchakovski, Sergei A. Voloshin, ``Electromagnetic Field Evolution in Relativistic Heavy-Ion Collisions'', {\changeurlcolor{vividviolet}\href{http://journals.aps.org/prc/abstract/10.1103/PhysRevC.83.054911}{Phys. Rev. C \textbf{83} (2011) 054911}}, \href{https://arxiv.org/abs/1103.4239}{[arXiv:1103.4239 [nucl-th]]}.

\bibitem{kn:bzdak}
Adam Bzdak, Vladimir Skokov, ``Event-by-Event Fluctuations of Magnetic and Electric Fields in Heavy Ion Collisions'', {\changeurlcolor{vividviolet}\href{http://www.sciencedirect.com/science/article/pii/S037026931200216X}{Phys. Lett. B \textbf{710} (2012) 171}}, \href{https://arxiv.org/abs/1111.1949}{[arXiv:1111.1949 [hep-ph]]}.

\bibitem{kn:shipu}
Victor Roy, Shi Pu, ``Event-by-Event Distribution of Magnetic Field Energy Over Initial Fluid Energy Density in $\sqrt{s_{NN}}=$ 200 GeV Au-Au Collisions'',{\changeurlcolor{vividviolet}
\href{http://journals.aps.org/prc/abstract/10.1103/PhysRevC.92.064902}{Phys. Rev. C \textbf{92} (2015) 064902}}, \href{https://arxiv.org/abs/1508.03761}{[arXiv:1508.03761 [nucl-th]]}.

\bibitem{kn:holliday}
Robert Holliday, Kirill Tuchin,
``Classical Electromagnetic Fields from Quantum Sources in Heavy-Ion Collisions'', \href{https://arxiv.org/abs/1604.04572}{[arXiv:1604.04572 [hep-ph]]}.

\bibitem{kn:liang}
Zuo-Tang Liang, Xin-Nian Wang, ``Globally Polarized Quark-Gluon Plasma in Non-Central A+A Collisions'',
{\changeurlcolor{vividviolet}\href{http://journals.aps.org/prl/abstract/10.1103/PhysRevLett.94.102301}{Phys. Rev. Lett. \textbf{94} (2005) 102301}, \href{http://journals.aps.org/prl/abstract/10.1103/PhysRevLett.96.039901}{Erratum-ibid. \textbf{96} (2006) 039901}}, \href{https://arxiv.org/abs/nucl-th/0410079}{[arXiv:nucl-th/0410079]}.

\bibitem{kn:bec}
Francesco Becattini, Fulvio Piccinini, Jose R. Rizzo, ``Angular Momentum Conservation in Heavy Ion Collisions at Very High Energy, {\changeurlcolor{vividviolet}\href{http://journals.aps.org/prc/abstract/10.1103/PhysRevC.77.024906}{Phys. Rev. C \textbf{77} (2008) 024906}},
\href{https://arxiv.org/abs/0711.1253}{[arXiv:0711.1253 [nucl-th]]}.

\bibitem{kn:huang}
Xu-Guang Huang, Pasi Huovinen, Xin-Nian Wang, ``Quark Polarization in a Viscous Quark-Gluon Plasma'',
{\changeurlcolor{vividviolet}\href{http://journals.aps.org/prc/abstract/10.1103/PhysRevC.84.054910}{Phys. Rev. C \textbf{84} (2011) 054910}}, \href{https://arxiv.org/abs/1108.5649}{[arXiv:1108.5649 [nucl-th]]}.

\bibitem{kn:KelvinHelm}
L\'aszl\'o P. Csernai, Daniel D. Strottman, Cs. Anderlik, ``Kelvin-Helmholz Instability in High Energy Heavy Ion Collisions'', {\changeurlcolor{vividviolet}\href{http://journals.aps.org/prc/abstract/10.1103/PhysRevC.85.054903}{Phys. Rev. C \textbf{85} (2012) 054901}},
\href{https://arxiv.org/abs/1112.4287}{[arXiv:1112.4287 [nucl-th]]}.

\bibitem{kn:viscous}
Dujuan Wang, Zolt\'an N\'eda, L\'aszl\'o P. Csernai, ``Viscous Potential Flow Analysis of Peripheral Heavy Ion Collisions'',
{\changeurlcolor{vividviolet}\href{http://journals.aps.org/prc/abstract/10.1103/PhysRevC.87.024908}{Phys. Rev. C \textbf{87} (2013) 024908}}, \href{https://arxiv.org/abs/1302.1691}{[arXiv:1302.1691 [nucl-th]]}.

\bibitem{kn:csernairecent1}
Sindre Velle, L\'aszl\'o P. Csernai, ``Differential Hanbury-Brown-Twiss for an Exact Hydrodynamic Model with Rotation'', {\changeurlcolor{vividviolet}\href{http://journals.aps.org/prc/abstract/10.1103/PhysRevC.92.024905}{Phys. Rev. C \textbf{92} (2015) 2, 024905}}, \href{https://arxiv.org/abs/1508.01884}{[arXiv:1508.01884 [nucl-th]]}.

\bibitem{kn:csernairecent2}
Sindre Velle, Sharareh Mehrabi Pari, L\'aszl\'o P. Csernai, ``Interferometry for Rotating Sources'', {\changeurlcolor{vividviolet}\href{http://www.sciencedirect.com/science/article/pii/S0370269316301125}{Phys. Lett. B \textbf{757} (2016) 501}}, \href{https://arxiv.org/abs/1508.04017}{[arXiv:1508.04017 [nucl-th]]}.

\bibitem{kn:nagy}
Tam\'as Cs\"org\H{o}, Marton I. Nagy, I. F. Barna, ``Observables and Initial Conditions for Rotating and Expanding Fireballs with Spheroidal Symmetry'', {\changeurlcolor{vividviolet}\href{http://journals.aps.org/prc/abstract/10.1103/PhysRevC.93.024916}{Phys. Rev. C \textbf{93} (2016) 024916}}, \href{https://arxiv.org/abs/1511.02593}{[arXiv:1511.02593 [nucl-th]]}.

\bibitem{kn:nacs}
Marton I. Nagy, Tam\'as Cs\"org\H{o}, ``An Analytic Hydrodynamical Model of Rotating 3D Expansion in Heavy-Ion Collisions'', \href{https://arxiv.org/abs/1512.00888}{[arXiv:1512.00888 [nucl-th]]}.


\bibitem{kn:yin}
Yin Jiang, Zi-Wei Lin, Jinfeng Liao, ``Rotating Quark-Gluon Plasma in Relativistic Heavy Ion Collisions'', {\changeurlcolor{vividviolet}\href{http://journals.aps.org/prc/abstract/10.1103/PhysRevC.94.044910}{Phys. Rev. C \textbf{94} (2016) no.4, 044910}}, \href{https://arxiv.org/abs/1602.06580}{[arXiv:1602.06580 [hep-ph]]}.

\bibitem{kn:deng}
Wei-Tian Deng, Xu-Guang Huang, ``Vorticity in Heavy-Ion Collisions'', {\changeurlcolor{vividviolet}\href{http://journals.aps.org/prc/abstract/10.1103/PhysRevC.93.064907}{Phys. Rev. C \textbf{93} (2016) 064907}}, \href{https://arxiv.org/abs/1603.06117}{[arXiv:1603.06117 [nucl-th]]}.

\bibitem{kn:vortical}
Long-Gang Pang, Hannah Petersen, Qun Wang, Xin-Nian Wang,
``Vortical Fluid and $\Lambda$ Spin Correlations in High-Energy Heavy-Ion Collisions'', \href{https://arxiv.org/abs/1605.04024}{[arXiv:1605.04024 [hep-ph]]}.

\bibitem{kn:plebdem}
Jerzy F. Pleba\'nski, Marek Demia\'nski, ``Rotating, Charged, and Uniformly Accelerating Mass in General Relativity'', {\changeurlcolor{vividviolet}\href{http://www.sciencedirect.com/science/article/pii/0003491676902402}{Ann. Phys. \textbf{98} (1976) 98}}.

\bibitem{kn:grifpod}
Jerry B. Griffiths, Jiri Podolsky, ``A New Look at the Pleba\'nski-Demia\'nski Family of Solutions'', {\changeurlcolor{vividviolet}\href{http://www.worldscientific.com/doi/abs/10.1142/S0218271806007742}{Int. J. Mod. Phys. D \textbf{15} (2006) 335}}, \href{https://arxiv.org/abs/gr-qc/0511091}{[arXiv:gr-qc/0511091]}.

\bibitem{kn:shear}
Brett McInnes, Edward Teo, ``Generalised Planar Black Holes and the Holography of Hydrodynamic Shear'', {\changeurlcolor{vividviolet}\href{http://www.sciencedirect.com/science/article/pii/S0550321313005774}{Nucl. Phys. B \textbf{878C} (2014) 186}}, \href{https://arxiv.org/abs/1309.2054}{[arXiv:1309.2054 [hep-th]]}.

\bibitem{kn:denghuang}
Wei-Tian Deng, Xu-Guang Huang, ``Event-by-Event Generation of Electromagnetic Fields in Heavy-Ion Collisions'', {\changeurlcolor{vividviolet}\href{http://journals.aps.org/prc/abstract/10.1103/PhysRevC.85.044907}{Phys. Rev. C \textbf{85} (2012) 044907}}, \href{https://arxiv.org/abs/1201.5108}{[arXiv:1201.5108 [nucl-th]]}.

\bibitem{kn:gergend}
Long-Gang Pang, Gergely Endrodi, Hannah Petersen, ``Magnetic Field-Induced Squeezing Effect at RHIC and at the LHC'', {\changeurlcolor{vividviolet}\href{http://journals.aps.org/prc/abstract/10.1103/PhysRevC.93.044919}{Phys. Rev. C \textbf{93} (2016) 044919}}, \href{https://arxiv.org/abs/1602.06176}{[arXiv:1602.06176 [nucl-th]]}.

\bibitem{kn:noodles}
Robert W. Moerman, William A. Horowitz, ``A Semi-Classical Recipe for Wobbly Limp Noodles in Partonic Soup'', \href{https://arxiv.org/abs/1605.09285}{[arXiv:1605.09285 [hep-th]]}.

\bibitem{kn:quench}
Carlota Andr\'es, N\'estor Armesto, Matthew Luzum, Carlos A. Salgado, Pía Zurita,
``Energy Versus Centrality Dependence of the Jet Quenching Parameter $\hat{q}$ at RHIC and LHC: A New Puzzle?'', 	 {\changeurlcolor{vividviolet}\href{http://link.springer.com/article/10.1140\%2Fepjc\%2Fs10052-016-4320-5}{Eur. Phys. J. C \textbf{76} (2016) 475}}, \href{https://arxiv.org/abs/1606.04837}{[arXiv:1606.04837 [hep-ph]]}.

\bibitem{kn:abdalla}
Shao-Jun Zhang, Elcio Abdalla, ``Holographic Thermalization in Charged Dilaton Anti-de Sitter Spacetime'', {\changeurlcolor{vividviolet}\href{http://www.sciencedirect.com/science/article/pii/S0550321315001650}{Nucl. Phys. B \textbf{896} (2015) 569}}, \href{https://arxiv.org/abs/1503.07700}{[arXiv:1503.07700 [hep-th]]}.

\bibitem{kn:gout}
Blaise Gout\'eraux, Elias Kiritsis, ``Generalized Holographic Quantum Criticality at Finite Density'', {\changeurlcolor{vividviolet}\href{http://link.springer.com/article/10.1007\%2FJHEP12\%282011\%29036}{JHEP \textbf{1112} (2011) 036}}, \href{https://arxiv.org/abs/1107.2116}{[arXiv:1107.2116 [hep-th]]}.

\bibitem{kn:gz}
Chang-Jun Gao, Shuang-Nan Zhang, ``Topological Black Holes in Dilaton Gravity Theory'', {\changeurlcolor{vividviolet}\href{http://www.sciencedirect.com/science/article/pii/S0370269305003710}{Phys. Lett. B \textbf{612} (2005) 127}}.

\bibitem{kn:ong}
Yen Chin Ong, Pisin Chen, ``Stringy Stability of Charged Dilaton Black Holes with Flat Event Horizon'', {\changeurlcolor{vividviolet}\href{http://link.springer.com/article/10.1007\%2FJHEP08\%282012\%29079}{JHEP \textbf{1208} (2012) 079}, \href{http://link.springer.com/article/10.1007\%2FJHEP01\%282015\%29083}{Erratum-ibid.\textbf{1208} (2012) 079}}, \href{https://arxiv.org/abs/1205.4398}{[arXiv:1205.4398 [hep-th]]}.

\bibitem{kn:myers}
Robert C. Myers, Miguel F. Paulos, Aninda Sinha,
``Holographic Hydrodynamics with a Chemical Potential', {\changeurlcolor{vividviolet}\href{http://iopscience.iop.org/article/10.1088/1126-6708/2009/06/006/meta}{JHEP \textbf{0906} (2009) 006}}, \href{https://arxiv.org/abs/0903.2834}{[arXiv:0903.2834]}.

\bibitem{kn:hartkov}
Sean A. Hartnoll, Pavel Kovtun, ``Hall Conductivity from Dyonic Black Holes'', {\changeurlcolor{vividviolet}\href{http://journals.aps.org/prd/abstract/10.1103/PhysRevD.76.066001}{Phys. Rev. D \textbf{76} (2007) 066001}}, \href{https://arxiv.org/abs/0704.1160}{[arXiv:0704.1160 [hep-th]]}.

\bibitem{1101.5776}
Yen Chin Ong, ``Stringy Stability of Dilaton Black Holes in 5-Dimensional Anti-de Sitter Space'', Proceedings of the Conference in Honor of Murray Gell-Mann's 80th Birthday, p.583-590, World Scientific (2010) Singapore, \href{https://arxiv.org/abs/1101.5776}{[arXiv:1101.5776 [hep-th]]}.

\bibitem{GHS}
David Garfinkle, Gary T. Horowitz and Andrew Strominger, ``Charged Black Holes in String Theory'', {\changeurlcolor{vividviolet}\href{http://journals.aps.org/prd/abstract/10.1103/PhysRevD.43.3140}{Phys. Rev. D, \textbf{43} (1991) 31403143}}, {\changeurlcolor{vividviolet}\href{http://journals.aps.org/prd/abstract/10.1103/PhysRevD.45.3888}{Erratum-ibid.\textbf{45} (1992) 3888}}.

\bibitem{G}
Gary W. Gibbons, ``Antigravitating Black Hole Solutions with Scalar Hair in N = 4 Supergravity'', {\changeurlcolor{vividviolet}\href{http://www.sciencedirect.com/science/article/pii/0550321382901705}{Nucl. Phys. B \textbf{207} (1982) 337}}.

\bibitem{GM}
Gary W. Gibbons, Kei-ichi Maeda, ``Black Holes and Membranes in Higher Dimensional Theories with Dilaton Fields'', {\changeurlcolor{vividviolet}\href{http://www.sciencedirect.com/science/article/pii/0550321388900065}{Nucl. Phys. B \textbf{298} (1988) 741}}.

\bibitem{9210119}
Gary T. Horowitz, ``The Dark Side of String Theory: Black Holes and Black Strings'',  Trieste 1992 Proceedings: String Theory and Quantum Gravity \textbf{55} (1992) 99, \href{https://arxiv.org/abs/hep-th/9210119}{[arXiv:hep-th/9210119]}.


\bibitem{kn:barb}
Barbara Betz, Miklos Gyulassy,
``Azimuthal Jet Tomography of Quark Gluon Plasmas at RHIC and LHC'', \href{https://arxiv.org/abs/1305.6458}{[arXiv:1305.6458 [nucl-th]]}.


\bibitem{kn:ficnar1}
 Andrej Ficnar, Steven S. Gubser, Miklos Gyulassy, ``Shooting String Holography of Jet Quenching at RHIC and LHC'', {\changeurlcolor{vividviolet}\href{http://www.sciencedirect.com/science/article/pii/S0370269314007394?np=y}{Phys. Lett. B \textbf{738} (2014) 464}}, \href{https://arxiv.org/abs/1311.6160}{[arXiv:1311.6160 [hep-ph]]}.

\bibitem{kn:ficnar2}
Andrej Ficnar, Steven S. Gubser, Miklos Gyulassy, ``Holographic Light Quark Jet Quenching at RHIC and LHC via the Shooting Strings'', {\changeurlcolor{vividviolet}\href{http://www.sciencedirect.com/science/article/pii/S0375947414002188}{Nucl. Phys. A \textbf{932} (2014) 264}},
\href{https://arxiv.org/abs/1404.0935}{[arXiv:1404.0935 [hep-ph].]}

\bibitem{kn:SppC}
Ning-bo Chang et al., ``Physics Perspectives of Heavy-Ion Collisions at Very High Energy'', {\changeurlcolor{vividviolet}\href{http://link.springer.com/article/10.1007\%2Fs11433-015-5778-0}{Sci. China Phys. Mech. Astron. \textbf{59} (2016) 621001}}, \href{https://arxiv.org/abs/1510.05754}{[arXiv:1510.05754 [nucl-th]]}.

\bibitem{kn:FCC1}
N\'estor Armesto, Andrea Dainese, David d'Enterria, Silvia Masciocchi, Christof Roland, Carlos A. Salgado, Marco van Leeuwen, Urs A. Wiedemann,
``Nuclear Collisions at the Future Circular Collider'', \href{https://arxiv.org/abs/1601.02963}{[arXiv:1601.02963 [hep-ph]]}.

\bibitem{kn:FCC2}
Andrea Dainese, Urs A. Wiedemann et al., ``Heavy Ions at the Future Circular Collider'', \href{https://arxiv.org/abs/1605.01389}{[arXiv:1605.01389 [hep-ph]]}.

\bibitem{kn:STAR}
Sabita Das (for the STAR collaboration), ``Chemical Freeze-out Parameters in Beam Energy Scan Program of STAR at RHIC'', {\changeurlcolor{vividviolet}\href{http://www.epj-conferences.org/articles/epjconf/abs/2015/09/epjconf_ismd2015_10003/epjconf_ismd2015_10003.html}{EPJ Web of Conf. \textbf{90} (2015) 10003}}, \href{https://arxiv.org/abs/1412.0350}{[arXiv:1412.0350 [nucl-ex]]}.

\bibitem{kn:shine}
Michael Unger (for the NA61/SHINE Collaboration), ``Results from NA61/SHINE'', {\changeurlcolor{vividviolet}\href{http://www.epj-conferences.org/articles/epjconf/abs/2013/13/epjconf_isvh2012_01009/epjconf_isvh2012_01009.html}{EPJ Web Conf. \textbf{52} (2013) 01009}}, \href{https://arxiv.org/abs/1305.5281}{[arXiv:1305.5281 [nucl-ex]]}.

\bibitem{kn:nica}
V. D. Kekelidze, A. D. Kovalenko, I. N. Meshkov, A. S. Sorin, G. V. Trubnikov, ``NICA at JINR: New Prospects for Exploration of Quark-Gluon Matter'',
{\changeurlcolor{vividviolet}\href{http://link.springer.com/article/10.1134/S1063778812050122}{[Phys. of Atomic Nuclei \textbf{75} (2012) 542]}}.

\bibitem{kn:fair}
Marcus Bleicher, Marlene Nahrgang, Jan Steinheimer, Pedro Bicudo, ``Physics Prospects at FAIR'', {\changeurlcolor{vividviolet}\href{http://www.actaphys.uj.edu.pl/vol43/abs/v43p0731}{Acta Phys. Polon. B \textbf{43} (2012) 731}}, \href{https://arxiv.org/abs/1112.5286}{[arXiv:1112.5286 [hep-ph]]}.

\bibitem{kn:BEAM}
Yasuyuki Akiba et al., ``The Hot QCD White Paper: Exploring the Phases of QCD at RHIC and the LHC'', \href{https://arxiv.org/abs/1502.02730}{[arXiv:1502.02730 [nucl-ex]]}.


\bibitem{kn:luo}
Xiaofeng Luo, ``Exploring the QCD Phase Structure with Beam Energy Scan in Heavy-Ion Collisions'', {\changeurlcolor{vividviolet}\href{http://www.sciencedirect.com/science/article/pii/S0375947416300112}{Nucl. Phys. A \textbf{956} (2016) 75}}, \href{https://arxiv.org/abs/1512.09215}{[arXiv:1512.09215 [nucl-ex]]}.

\bibitem{kn:86}
Brett McInnes, ``Inverse Magnetic/Shear Catalysis'', {\changeurlcolor{vividviolet}\href{http://www.sciencedirect.com/science/article/pii/S0550321316000717}{Nucl. Phys. B \textbf{906} (2016) 40}}, \href{https://arxiv.org/abs/1511.05293}{[arXiv:1511.05293 [hep-th]]}.

\bibitem{kn:87}
Brett McInnes, ``A Rotation/Magnetism Analogy for the Quark–Gluon Plasma'', {\changeurlcolor{vividviolet}\href{http://www.sciencedirect.com/science/article/pii/S055032131630219X}{Nucl. Phys. B \textbf{911} (2016) 173}}, \href{https://arxiv.org/abs/1604.03669}{[arXiv:1604.03669 [hep-th]]}.









\end{thebibliography}
\end{document}